\documentclass[preprint,12pt]{elsarticle}

\usepackage{amssymb}

\journal{Astroparticle Physics}

\begin{document}

\begin{frontmatter}



\title{Testing Lorentz Invariance with Neutrinos from Ultrahigh Energy Cosmic Ray Interactions}


\author{Sean T. Scully}
\ead{scullyst@jmu.edu}
\address{Department of Physics and Astronomy, James Madison University,Harrisonburg, VA 22807}

\author{Floyd W. Stecker}
\ead{Floyd.W.Stecker@nasa.gov}  
\address{NASA/Goddard Space Flight Center, Greenbelt, MD 20771}

\begin{abstract}
We have previously shown that a very small amount of Lorentz invariance violation
(LIV), which suppresses  photomeson  interactions  of ultrahigh  energy cosmic rays  (UHECRs) with 
cosmic background radiation (CBR) photons, can produce a spectrum of cosmic rays that is consistent 
with that currently observed by the Pierre Auger Observatory (PAO) and HiRes experiments. Here, we 
calculate the corresponding flux of high energy neutrinos generated by the propagation of UHECR protons 
through the CBR in the presence of LIV. We find that LIV produces a reduction in the flux of the 
highest energy neutrinos and a reduction in the energy of the peak of the neutrino energy 
flux spectrum, both depending on the strength of the LIV. Thus, observations of the UHE neutrino 
spectrum provide a clear test for the existence and amount of LIV at the highest energies. We further 
discuss the ability of current and future proposed detectors make such observations.
\end{abstract}


\end{frontmatter}

\newpage



\section{Introduction}
\label{intro}
Ultrahigh energy cosmic rays and neutrinos are of interest as possible probes of new physics \citep
{ste05}. In particular, some quantum gravity  models predict that Lorentz invariance may be weakly 
broken at the very high energies, leading to potentially observable consequences. The possibility of 
using ultrahigh energy cosmic rays (UHECRs) to probe for a small violation of Lorentz invariance was 
suggested over a decade ago \citep{col99}.  Indeed, a detailed analysis of the effects of LIV on the 
UHECR spectrum has yielded the tightest constraint on LIV to date \citep{ste09}.  Shortly after the 
discovery of the CBR it was pointed out that photomeson interactions of UHECRs with photons of the 
cosmic background radiation (CBR) would result in a sharp steepening of their spectrum above E $\sim$ 50 EeV now known as the "GZK effect"~\citep{gre66, zat66}. However, even a very small amount of LIV will kinematically inhibit some of these interactions. It has been previously shown that a possible signature of LIV in the UHECR spectrum would be a  recovery of the cosmic ray spectrum at energies greater than $\sim$ 200 EeV (\citep{ste09, scu09}). 

Given the current state of the UHECR observational data as reported by HiRes \citep{bel09, sok09} and Auger \citep{abr10}, it is possible to constrain LIV, but not to rule it out.  However, unlike cosmic-ray baryons and photons, ultrahigh energy neutrinos do not suffer significant energy losses
over cosmological distances.   Studies of UHE neutrinos with both ground-based and space-based detectors could provide a new and less ambiguous test of LIV. This is because
they are a guaranteed byproduct of the photomeson interactions of UHECRs with the CBR followed by 
subsequent pion decay \citep{ber70, ste73, ste79, eng01}.

In this paper, we further consider the observational implications of the
effect of a very small amount of LIV, {\it viz.} the suppression of photomeson production on the 
resulting UHE neutrino spectrum.  In our previous work \citep{ste09, scu09} we undertook a detailed 
calculation of the modification of the UHECR spectrum caused  by LIV using the formalism of reference \cite{col99} and the  kinematical approach  originally  developed in reference \cite{alf03}. We employ the same techniques used in references \cite{ste09} and ~\cite{scu09} to determine the resulting photomeson neutrino spectrum. We again consider here the case where the primary UHECRs are protons. 

ANITA II \cite{gor10} has placed an upper limit on the neutrino flux for energies greater than $>10^{18}$ eV. IceCube is nearing completion and will be sensitive to neutrinos of energies up to $\sim 10^{17}$eV.  Proposed future ground-based and space-based neutrino detectors could be capable of detecting and  studying photomeson neutrinos.  We will discuss the ability of such detectors to constrain LIV or observe its effect.  

\section{LIV and the Spectrum of UHECRs}
\label{liv}

We now extend the calculation of reference \citep{ste09} to determine the photomeson neutrino fluxes.\footnote{We use the usual convention $c = 1$.}~A full description of the LIV formalism we use is given in references \citep{col99} and \citep{ste09}. We summarize the salient points here.  The free particle Lagrangian is modified by the inclusion of a leading order perturbative, Lorentz violating term. This term leads to the modified free particle dispersion relations 

\begin{equation}
E^2 ~ = ~ \vec{p} \ ^{2} + m^{2} + 2\delta \vec{p} \ ^2.
\label{dispersion}
\end{equation}

These relations can be put in the standard form  
\begin{equation}
E^2 ~ =~  \vec{p \ }{^2}c_{MAV}^2 + m^{2} c_{MAV}^4,
\end{equation} 
by shifting the renormalized mass by the small amount $m \rightarrow m/(1+2\delta)$ and shifting the 
velocity by the amount 

\begin{equation}
c_{MAV} = \sqrt{(1 +  2\delta)} \simeq 1 + \delta 
\end{equation} 
where $c_{MAV}$ is 
identified as the maximum attainable velocity of the free particle in the reference frame of the CBR.  Using this formalism, different 
particles can have different maximum attainable velocities (MAVs) that can all be different from $1$ as 
well as different from one another.  Hereafter, we denote the MAV of a
particle of type $i$  by $c_{i}$  and the difference 
\begin{equation}
c_{i}  - c_{j} ~ ~ \equiv ~ \delta_{ij}
\label{deltadef}
\end{equation}

These modified dispersion relations are then applied to the kinematical relations governing the dominant single meson photomeson interaction:
\begin{equation}
p + \gamma \rightarrow N + \pi.
\end{equation}
From equations (\ref{dispersion}) and (\ref{deltadef}), a dispersion relation can be constructed for a 
particle $a$
\begin{equation}
E^2=p^2+2\delta _a p^2 +{m_a}^2
\label{particledisp}
\end{equation}
where $\delta _a$  is the difference between the  MAV for the particle
{\it a} and  the speed  of light in the low momentum limit, $c = 1$.

In order to modify the effect of photomeson production on the UHECR spectrum above the GZK energy, $
\delta_{\pi p} > 0$ as shown in reference \citep{col99}.
The condition for photomeson interactions to take place is

\begin{equation}
\delta_{\pi p} \le 3.23 \times 10^{-24} (\omega/\omega_0)^2.
\label{CG}
\end{equation}

\noindent where $\omega$ is the energy of the CBR photon and $\omega_0 \equiv  kT_{CBR} =  2.35 \times  
10^{-4}$  eV with $T_{CBR} = 2.725\pm 0.02$ K \cite{col99, ste09}. 

\begin{figure}
\begin{center}
\includegraphics[height=2.6in]{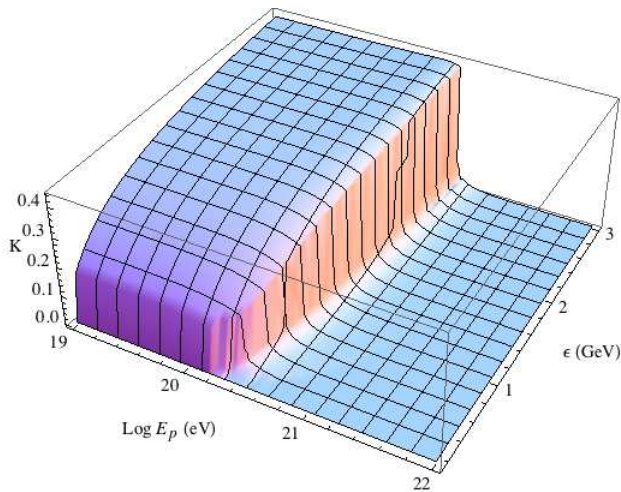}
\end{center}
\caption{The  calculated  proton inelasticity  modified  by LIV  for
$\delta_{\pi p} =  3 \times 10^{-23}$ as  a function of CBR  photon energy and
proton energy (from reference ~\citep{ste09}).}
\label{inelasticity}
\end{figure}

If LIV occurs and $\delta_{\pi p} > 0$, photomeson production can only take place for interactions of  CBR photons with energies large enough to satisfy equation (\ref{CG}). This condition implies  that while photomeson  interactions leading to GZK suppression can occur for ``lower energy'' UHE protons interacting with higher energy CBR photons on the Wien tail of the spectrum, other interactions involving higher energy  protons and photons with smaller values  of  $\omega$ will  be  forbidden \cite{ste09}.   Thus,  the observed  UHECR spectrum may  exhibit the characteristics of GZK  suppression near the normal GZK threshold, but the UHECR spectrum can ``recover'' at higher energies  owing to  the  possibility that  photomeson interactions  at higher  proton energies  may be  forbidden ~\cite{ste09,scu09,mac09,bix09}.
Even a small violation of Lorentz invariance changes the inelasticity of the interaction, ({\it i.e.}, the amount of energy transferred from the incident proton to the created pion). This is the key to understanding the effect of LIV on 
photomeson production.  With an increase in proton energy, the range of kinematically allowed 
angles of interaction between it and the photon becomes  more restricted, thus reducing the phase 
space and, in turn, the total inelasticity. Figure \ref{inelasticity}, reproduced from reference \citep{ste09}, shows the calculated 
inelasticity modified by LIV for a value of $\delta_{\pi p} =  3 \times 10^{-23}$ as a function of both 
CBR photon energy and incident proton energy.    Other choices for $\delta_{\pi p}$ yield similar plots but change the energy at which LIV effects become significant.
 The inelasticity precipitously drops above a certain 
energy because the LIV term in the pion rest energy from equation (\ref{particledisp}) becomes comparable to $m_{\pi}$.   

The  proton energy loss rate by photomeson production is given by
\begin{equation}
{{1}\over{E}}{{dE}\over{dt}}     =     -    {{\omega_{0}c}\over{2\pi^2
\gamma^2}\hbar^3c^3} \int\limits_\eta^\infty  d\epsilon ~ \epsilon ~
\sigma(\epsilon)                                            K(\epsilon)
\ln[1-e^{-\epsilon/2\gamma\omega_{0}}]
\label{pionproduction}
\end{equation}
where $\epsilon$ is the photon energy in the center of mass system,   $K(\epsilon)$ is the  modified 
inelasticity calculated from the kinematics, and
$\sigma(\epsilon)$ is  the total $\gamma$-p cross section. 
The lower limit of the integration,$\eta$, is the photon threshold energy for the interaction in the 
center of mass frame.

\begin{figure}
\begin{center}
\includegraphics[height=2.8in]{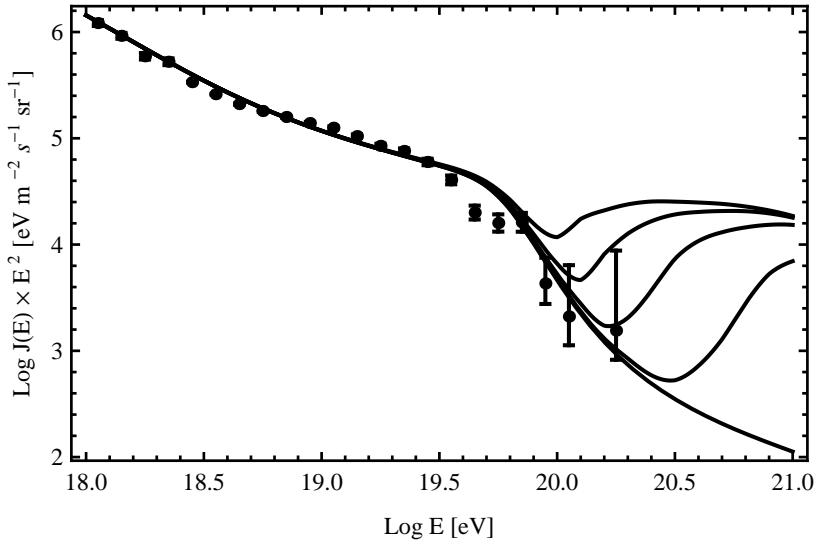}
\end{center}
\caption{Comparison of the PAO data \cite{abr10} with calculated spectra for various 
values of $\delta_{\pi p}.$ From top to bottom, the curves give the predicted 
spectra for 
$\delta_{\pi p} = 1 \times 10^{-22}, 6 \times 10^{-23}, 3 \times 10^{-23} , 1 \times 10^{-23},
0$ (no Lorentz violation).}
\label{Auger}
\end{figure}

As in reference \citep{ste09}, we assume that the source spectrum of UHE protons can be approximated 
over a limited energy range by a power-law that fits the UHECR data below 60 EeV. This spectrum is then 
of the form $A(z)E_i^{-\Gamma}$ where $E_{i}$ is the energy of the proton.  The UHECRs suffer energy 
losses from pair and pion production through interactions with the CBR and also cosmological 
redshifting.  The energy losses from pion production are determined according to equation (\ref
{pionproduction}).  The pair-production  loss  rate comes from \citep{blu70}.  In order to determine 
redshift loses, a flat $\Lambda$CDM  universe with  a Hubble  constant of H$_0$ = 70 km s$^{-1}$ Mpc$^
{-1}$ is assumed, taking $\Omega_{\Lambda}$ = 0.7 and $\Omega_{m}$  = 0.3. The source  evolution is 
additionally assumed $\propto (1+z)^{\zeta}$
with $\zeta$ = 3.6, out to  a maximum redshift of 2.5 which tracks the star formation rate. This value is a mean between the fast evolution and baseline models used in reference \cite{sms06}. (See also
references \cite{lef05,hop06}.) The 
spectrum of UHECRs on Earth can then be determined from
\begin{eqnarray} 
J(E) = 
{{3cA(0)}\over{8\pi H_{0}}} E^{-\Gamma}\int_{0}^{z_{max}}{{(1+z)^{(\zeta-1)}}\over{\sqrt{\Omega_{m}
(1+z)^3 +\Omega_{\Lambda}}}}
\left({E_i\over{E}}\right)^{-\Gamma}{{dE_i}\over{dE}}dz.
\end{eqnarray}
where A(0) is determined by fitting our final calculated spectrum to the observational UHECR data 
assuming $\Gamma=2.55$, which  is
consistent with the spectrum derived by the Pierre Auger Observatory (PAO) collaboration below 60 EeV.\footnote{We have chosen a maximum UHECR energy of $3 \times 10^{21}$ eV. However, our results are insensitive to this value because they are determined by the LIV kinematics.}  
The results of this calculation for various choices of the parameter $\delta_{\pi p}$ are shown in 
Figure \ref{Auger} plotted along with the most recent results from PAO.

It can be seen from Figure \ref{Auger} that a small amount of LIV can still preserve the GZK
suppression effect, but produces a "recovery" of the UHECR spectrum at higher energies. Since this
recovery is due to the virtual elimination of photomeson interactions at higher UHECR energies, 
it will also suppress the production of higher energy photomeson neutrinos.

\section{The Photomeson Neutrino Spectrum}
\label{neutrinos}

We now turn our attention to calculating the photomeson neutrino spectrum that would result from the 
UHECR calculation detailed in the previous section. We use the data on the cross
section for pion production compiled in reference \cite{arn02} and summarized in reference
\cite{ber07}. Near threshold the the total photomeson cross section is dominated by the emission of 
single pions.  The most significant channel to consider involves the intermediate
production of the $\Delta$ resonance ~\cite{ste68}:

\begin{equation} 
p + \gamma \rightarrow \Delta \rightarrow N + \pi 
\end{equation}
This channel strongly dominates the photomeson production process near 
threshold. Since the UHECR flux falls steeply with energy, it follows that the bulk of the pions 
leading to the production of neutrinos will be produced close to the threshold.   

For a proton interacting with the CBR, a pion and a nucleon are produced. The outgoing nucleon has probability of 2/3 to be a proton and 1/3 probability to be a neutron from isospin considerations.  Should the resulting nucleon be a neutron, then the resulting pion is a $\pi ^{+}$.  Thus approximately twice the number of neutral pions are produced relative to charged pions from resonant pion production.  However direct pion production, which accounts for about 20\% of the total cross section, produces charged pions almost exclusively meaning that all told, approximately equal numbers of neutral and charged pions are produced around threshold.   Neutral pions decay into photons so we need only consider the charged pions for neutrino production. Three neutrinos of roughly equal energy result from the decay chain of the $\pi ^{+} \rightarrow \mu^{+}\nu_{\mu} \rightarrow e^{+}\bar{\nu{_\mu}}\nu_{e}$.  

It is straightforward to determine the neutrinos produced and their energies from the UHECRs.  We follow closely the calculation of the neutrino flux as described in reference \citep{ste79} and references therein.  The key is to determine the amount of energy that is carried away by the pion.   This follows directly from the inelasticity and the incident proton energy calculated using equation (\ref{pionproduction}).   We assume that all of the sources have the same primary injection spectrum and distribution as detailed in section \ref{liv}.  We calculate the total neutrino flux by integrating over proton energy, photon energy, and redshift, assuming the standard $\Lambda$CDM
cosmology. We find that the shape of the neutrino spectrum we obtain when assuming Lorentz invariance is very similar to that obtained from the more detailed Monte Carlo calculations that include all the relevant baryonic resonances and possible meson and multi-pion production channels  (See, {\it e.g.}, reference \cite{eng01}).           

The effect of LIV on the photomeson neutrino production is again manifested through the modification of the inelasticity of the interaction, since this determines the amount of energy that is carried away by the pion and therefore the resultant neutrino energy. The biggest impact of including LIV is to suppress the production of the higher energy photopions and therefore the resulting higher energy neutrinos. 

\begin{figure}
\begin{center}
\includegraphics[height=2.8in]{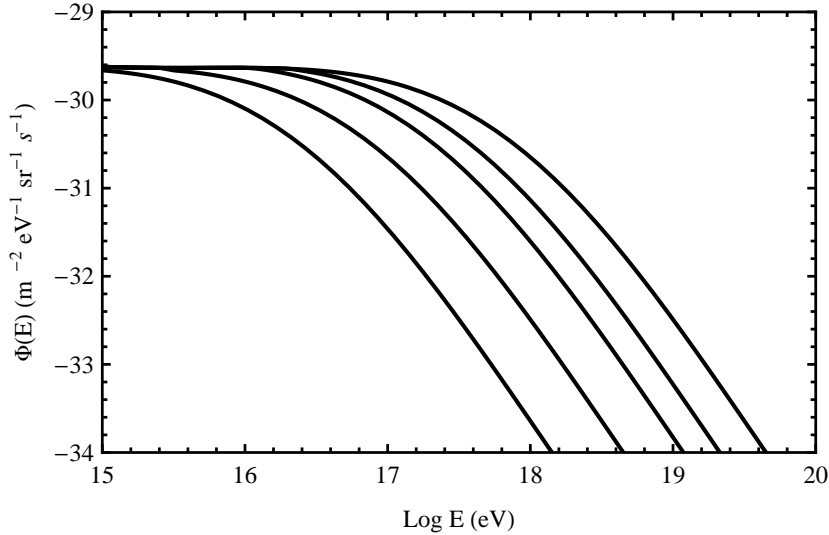}
\end{center}
\caption{Neutrino fluxes (of all species) corresponding to the UHECR models considered in Figure \ref{Auger}. From left to right, the curves give the predicted fluxes for $\delta_{\pi p} = 1 \times 10^{-22}, 6 \times 10^{-23}, 3 \times 10^{-23} , 1 \times 10^{-23}, 0$.}
\label{neutrinoflux}
\end{figure}

Figure \ref{neutrinoflux} shows the corresponding total neutrino flux (all species) for the same choices of $\delta_{\pi p}$ as the UHECR spectra presented in figure \ref{Auger}.  As expected, increasing $\delta_{\pi p}$ leads to a decreased flux of higher energy photomeson neutrinos as the interactions involving higher energy UHECRs are suppressed \cite{ste09}.  It is also evident that the peak energy of the neutrino energy flux spectrum (EFS), $E\Phi(E)$, shifts to lower energies with increasing $\delta_{\pi p}$.    
We note that a similar effect can be produced on the photomeson neutrino spectrum from another
possible mechanism involving LIV, namely "neutrino splitting" ~\cite{mat10}. With LIV, the decay of one neutrino into three neutrinos can be kinematically allowed. This effect may also produce a decreased flux at the 
high energy end of the neutrino spectrum. The feature that distinguishes between the two possible
LIV effects is that neutrino splitting results in an increase in the flux of lower energy
photomeson neutrinos.

\section{Considerations of UHECR Composition}

Throughout this  paper, we have  made the assumption that  the highest energy cosmic rays, {\it i.e.}, those above 100 EeV, are protons.  The composition  of these  primary  particles is  presently unknown.   The highest  energy events  for which  composition measurements  have been attempted are in the range between  40 and 50 EeV, and the composition of these events is uncertain~\cite{be09},\cite{ung09},\cite{ulr09}.

We note  that in  the case  where the UHECRs  with total  energy above $\sim$100 EeV are  not protons, both the photomeson  threshold and the LIV effects are moved to  higher energies because (i) the threshold is dependent on $\gamma  \propto E/A$, where $A$ is  the atomic weight of the   UHECR   \cite{ste69},  and   (ii)   it   follows  from   equation (\ref{dispersion})  that  the LIV  effect  depends  on the  individual nucleon momentum. We also note that the neutrino spectrum at the high energy end is the same for the mixed composition case as in the pure proton case \cite{hoo05,ave05,all06}.

\section{Observational Prospects}
\label{observations}
Several experiments currently place upper limits on the photomeson neutrino flux.  The ANITA long 
duration balloon experiment launched in December of 2008 searched for  
electromagnetic cascades initiated by UHE neutrinos within the Antarctic ice shelf {\it via} the Askaryan effect. Their analysis 
yielded a model-independent 90\% CL limit on neutrino fluxes in the range of 10$^{18}$ -- 10$^{23}$ eV with a sensitivity capable of excluding several optimistic photomeson neutrino flux models
~\cite{gor10}. However, ANITA does not have sufficient sensitivity or energy range to distinguish a possible LIV effect on the neutrino spectrum. ANITA has an effective  threshold energy $\sim 10^{18}$ eV and its sensitivity about an order of magnitude too weak. IceCube is capable of detecting neutrinos $< 10^{17}$ eV, but its sensitivity is two orders of magnitude too weak to detect photomeson neutrinos~\cite{ahr04}.  
Proposed future space-borne missions such as the Extreme-Universe Space Observatory (EUSO) ~\cite
{san10}, super EUSO~\cite{pet09}, and Orbiting Wide-angle Light Detectors (OWL) ~\cite{ste04} would 
have much larger effective aperatures than presently available detectors. They would be capable of making accurate determinations of the energy, arrival direction, and composition of the cosmic-rays and the associated photomeson neutrinos using a target volume far greater than is possible from ground-based experiments presently in operation.  While such experiments could potentially provide the statistics necessary to observe the recovery of the UHECR spectrum at high energies, both EUSO and OWL as proposed would not achieve either the sensitivity or energies necessary to distinguish LIV effects in the neutrino spectrum.  

However, the proposed full Antarctic Ross Ice shelf ANtenna Neutrino Array (ARIANNA) would be capable 
of detecting photomeson neutrinos with a sensitivity an order of magnitude better than ANITA and other 
existing detectors. Like ANITA, ARIANNA exploits the Askaryan effect, {\it i.e.} the detection of 
coherent \v{C}erenkov emission at radio wavelengths produced in the ice shelf by neutrino induced cascades. Because the power of coherent radio emission grows as the square of the shower energy and therefore neutrino energy, ARIANNA is capable of detecting lower energy neutrinos than ANITA since the distances to balloon-borne detectors can be quite large while ARIANNA utilizes detectors situated directly on the ice shelf surface. ARIANNA is expected to observe $\sim$40 events per 6 months in the energy range of photomeson neutrinos with energies in excess of $\sim 10^{17}$ eV \citep{bar08}.  This lower energy threshold is crucial for searching for LIV effects.  As such, we shall restrict ourselves to the discussing the potential of ARIANNA for distinguishing the effect of LIV on the photomeson neutrino spectrum. We note that another 
proposed detector called IceRay would also make use of the Askaryan effect. IceRay would be placed at the location of IceCube and would also  be capable of detecting photomeson neutrinos
~\cite{all09}.

\begin{figure}
\begin{center}
\includegraphics[height=2.8in]{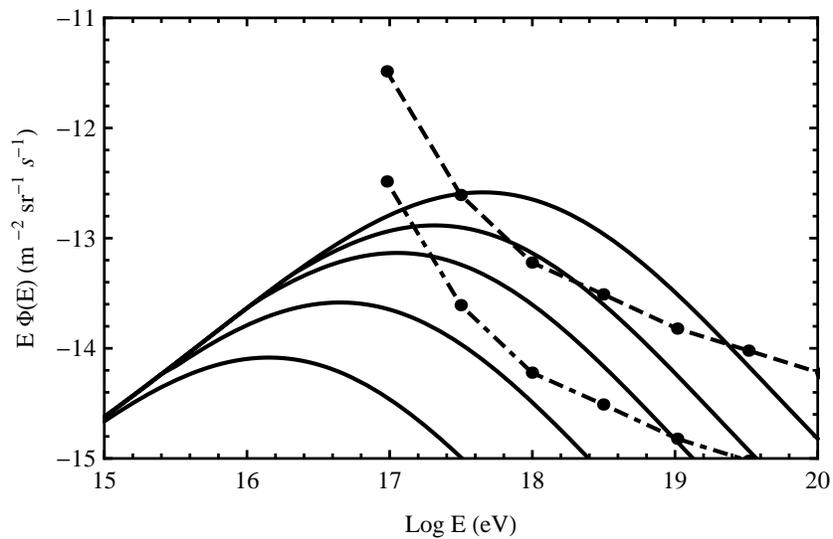}
\end{center}
\caption{All-flavor neutrino flux spectra (EFS) that correspond to the UHECR models considered in 
Figure \ref{Auger}. The six month sensitivity from the proposed ARIANNA array is shown as a dashed curve. The dot-dashed curve shows the sensitivity scaled to a 5 year exposure.  From bottom to top, the solid curves give the predicted spectra for $\delta_{\pi p} = 1 \times 10^{-22}, 6 \times 10^{-23}, 3 \times 10^{-23} , 1 \times 10^{-23}, 0$.}
\label{neutrinospec}
\end{figure}

ARIANNA's high event rate, combined with its low energy threshold can distinguish LIV effects if $\delta_{\pi p} \le 3.0 \times 10^{-23}$ with a 5 year exposure. We note that this limit is close to the upper limit indicated by the current PAO data \citep{ste09}.  Figure \ref{neutrinospec} shows our calculated fluxes along with the sensitivity of ARIANNA for exposure times of 6 months \citep{bar08} and 5 years.  Here we have plotted $E\Phi(E)$ for all neutrino flavors and $\nu$-$\bar{\nu}$ combinations compared with the proposed ARIANNA sensitivities to more clearly illustrate the threshold effect.   We note that ARIANNA is sensitive to all neutrino flavors since it would primarily detect the hadronic shower and not the outgoing lepton. 
In Figure \ref{neutrinospec}, we present our neutrino fluxes and experimental sensitivities as the total of all flavors and $\nu$-$\bar{\nu}$ combinations.      

It is clear from Figure \ref{neutrinospec} that at the lower energy threshold of $\sim 10^{17}$ eV for ARIANNA is very close to the peak energy in $E\Phi(E)$ for $\delta_{\pi p} = 3 \times 10^{-23}$ and would have sufficient sensitivity at $10^{17}$ eV to detect the expected neutrino flux provided it runs long enough to produce the desired sensitivity.  ARIANNA is therefore very promising in terms of its ability to distinguish the effect of an amount of LIV that is compatible with the current PAO results.  

\subsection{A Caveat}

We note that suggestions have been made to use the derivation of the extragalactic $\gamma$-ray 
background flux up to 100 GeV using data from the {\it Fermi} Gamma-ray Space Telescope~\cite{egb10} to 
place limits on the absolute value of the photomeson neutrino flux, assuming that the 100 GeV $\gamma$-
rays are produced by an electromagnetic cascade off the CBR initiated by UHE pion decay $\gamma$-rays 
produced along with the neutrinos \cite{ber10},\cite{ahl10}. Such a
constraint would lower the neutrino flux to almost an order of magnitude below the expected value
and would make it much more difficult to test Lorentz invariance with photomeson neutrinos.
These results follow earlier neutrino constraints ~\cite{sem04} obtained using various analyses of the 
EGRET-GRO results on the extragalactic $\gamma$-ray background~\cite{sre98},\cite{str04}.
However, the argument here is contingent on the assumption that the extragalactic magnetic field 
is so small that a UHE electromagnetic pair-production-Compton cascade 
leading to the production of $\gamma$-rays in the GeV energy range will not be cut off by synchrotron 
losses of the UHE electrons dominating over Compton losses \cite{wdo72, ste73}. At this point in time, 
the strength of the extragalactic B-field is only constrained to be within the range $\sim 3 \times 10^{-16}$ -- $\sim 3 \times 10^{-9}$ G.~\cite{kro08},\cite{ner10}.

In addition, we note that the {\it Fermi} spectrum is not the result of direct observation, but of 
analysis. It critically involves the subtraction of both galactic foreground $\gamma$-rays and, in the 
case of {\it Fermi} instrumental calibration, by Monte Carlo modeling~\cite{atw09}. These are non-negligible 
uncertainties. As an example of the uncertainties involved, we note the significant differences between 
the EGRET-GRO results ~\cite{sre98},\cite{str04} and the {\it Fermi} results ~\cite{egb10} on the extragalactic $\gamma$-ray background.

We also note that since both the photomeson production cross section and the CBR photon spectrum are
very well determined, and since the GZK cutoff effect is well documented~\cite{sok09},\cite{abr10}, a significant decrease in the predicted photomeson neutrino flux could either imply an unexpectedly small production of UHECRs or the existence of new physics. 

\newpage

\section{Conclusion}

With future improved data from PAO, tighter constraints can be placed on the amount of LIV allowed by the UHECR spectrum.  However even after a decade more of operation it seems unlikely that PAO would be able to determine the UHECR spectrum with adequate statistics at energies greater that 300 EeV, where the effect of LIV would be manifested.  While space-borne missions such as JEM-EUSO and OWL could provide the statistics necessary to observe the effect, these missions are currently only in the planning stage and are many years from being realized.  

We have shown here that additional information on LIV or its constraints can be obtained from studying the spectrum of photomeson neutrinos. 
By calculating the flux of high energy neutrinos generated by the propagation of UHECR protons 
through the CBR in the presence of LIV, we find that LIV produces a reduction in the flux of the 
highest energy photomeson neutrinos and a reduction in the energy of the peak of the neutrino energy 
flux spectrum with both effects increasing with the strength of the LIV. Thus, observations of the UHE neutrino spectrum could provide a clear test for the existence and amount of LIV that would be exhibited in the highest energy cosmic-ray interactions.

ARIANNA would have a sufficiently low threshold energy and the sensitivity necessary to determine the location of the energy peak in the photomeson neutrino EFS to further test LIV in the range  $\delta_{\pi p} \le 3 \times 10^{-23}$, consistent with the current limit indicated by the PAO data.  The amount of LIV or its nondetection has important consequences for Planck scale physics and quantum gravity theories.     

\section*{Acknowledgments}

We would like to thank Steve Barwick, Peter Gorham, and John Krizmanic for enlightening discussions and comments about the various present and potential future neutrino detectors.

\section*{References}




\bibliographystyle{elsarticle-num}
\bibliography{cr_bibtex}







\end{document}